\documentclass[aps,prd,superscriptaddress,showpacs,amsmath,amssymb]{revtex4}
\usepackage{graphicx,epsf}
\usepackage{amsfonts}
\usepackage{amssymb}
\usepackage[dvips]{color}
\definecolor{red}{rgb}{1,0,0}
\definecolor{green}{rgb}{0,1,0}

\newcommand{\be}{\begin{eqnarray}}
\newcommand{\ee}{\end{eqnarray}}
\newcommand{\bra}[1]{\mbox{$\langle\, #1 \mid$}}

\newcommand{\ket}[1]{\mbox{$\mid #1\,\rangle$}}

\begin{document}

\title{Resonant active-sterile neutrino mixing in the presence of matter potentials and altered dispersion relations}

\author{Sebastian Hollenberg}
\email{sebastian.hollenberg@uni-dortmund.de}
\author{Heinrich P\"as}
\email{heinrich.paes@uni-dortmund.de}
\affiliation{Fakult\"at f\"ur Physik, Technische Universit\"at Dortmund, D-44221 Dortmund, Germany}

\begin{abstract}
We investigate resonant active-sterile neutrino oscillations emergent from the interplay between matter, CPT- and Lorentz-violating coefficients with different energy dependences. Novel and experimentally 
accessible resonant neutrino oscillation phenomenologies are predicted and possible implications
for observed, but yet unexplained, neutrino oscillation anomalies such as LSND and MiniBooNE are discussed. 
\end{abstract}
\preprint{DO-TH-09/02}

\maketitle

\section{Introduction }

The data released by the MiniBooNE collaboration \cite{AguilarArevalo:2008rc} reveal a resonance-like excess of events in the low-energy neutrino channel, but do not show a deviation from the expected oscillation pattern in the antineutrino channel. The LSND collaboration \cite{Aguilar:2001ty} on the other hand observes an excess in the antineutrino channel.
In order to understand these yet unexplained anomalies neutrino oscillation scenarios with altered dispersion relations have been subject of discussion in the past.
Altered dispersion relations arise due to effective interactions of neutrinos with background fields and 
particles or are attributable to effects of new physics, e.g. CPT-violating extensions of the standard model 
\cite{Colladay:1996iz, Colladay:1998fq, Kostelecky:2003cr} or theories with additional spacetime dimensions \cite{ArkaniHamed:1998rs, Randall:1999ee}. The quantum mechanics of altered dispersion relations models can be understood as follows:
\par
The flavor-oscillation amplitude for a propagating neutrino state is
    \be
        A\left(\nu_{\alpha} \to \nu_{\beta}\right) = \bra{\nu_{\beta}}e^{-iHt}\ket{\nu_{\alpha}},
    \ee
where the part of the Hamiltonian $H$ proportional to the identity does not effect the flavor change and
can therefore be subtracted. The remainder can be written as $\delta(Ht)$ and under the assumption that the 
non-standard contributions are small we may expand $\delta(Ht) = t\delta H + H\delta t$. This leaves us with
    \be
        A\left(\nu_{\alpha} \to \nu_{\beta}\right) = \bra{\nu_{\beta}}e^{-i\left[t\delta H +
        H\delta t\right]}\ket{\nu_{\alpha}}.
    \ee
A non-vanishing second term in the exponential is unconventional and occurs if the travel times for neutrino 
states are not universal. Such a theory assigns different ``light-cones'' to different states and thereby
breaks Lorentz invariance \cite{Coleman:1998ti}. 
\par
Matter effects \cite{Wolfenstein:1977ue, Barger:1980tf, Mikheev:1986wj} in neutrino oscillations are just one specific example of neutrino
oscillations with altered dispersion relations. The emergent effective potential in the Schr\"odinger equation 
describing coherent elastic forward scattering of active neutrinos in medium discriminates between electron neutrinos which interact with matter via both charged and neutral currents and muon and tau neutrinos which only encounter neutral current interactions. Hypothetical $SU(2)$ singlet ``sterile'' neutrinos as considered in this work would not interact at all \cite{Maltoni:2007zf}. The additional contribution to the Hamiltonian varies linearly with the neutrino energy giving rise to resonant neutrino mixing in matter.
\par
Another example for altered dispersion relations are models with extra-dimensional shortcuts 
\cite{Pas:2005rb, Hollenberg:2009ws} of sterile neutrinos leading to a resonance in active-sterile neutrino mixing
in this model. 
Shortcuts of sterile neutrinos through an extra-dimensional bulk \cite{ArkaniHamed:1998rs, Randall:1999ee} give rise to an effective potential in the 
Schr\"odinger equation contributing to the sterile-sterile component of the Hamiltonian. The potential is 
parametrized in terms of the shortcut parameter, which contains information about the different 
travel times of the sterile neutrino propagating through the extra-dimensional bulk as well as on 
the brane. The effective potential for extra-dimensional shortcuts varies quadratically with the 
neutrino energy. Such extra-dimensional models break Lorentz invariance and hence are one specific subset of
CPT-violating extensions of the standard neutrino oscillation framework.
\par
Two problems with accommodating data in a scenario with altered dispersion relations in the sterile neutrino sector in vacuo, e.g. motivated via extra-dimensional shortcuts, have been noted. Firstly, the resonance energies for sterile 
neutrinos and antineutrinos are the same, which seems to be excluded by the latest MiniBooNE results \cite{AguilarArevalo:2009xn}. This calls for
a mechanism capable of realizing different resonance energies for neutrinos and antineutrinos. Another problem
is that the width of the resonance in shortcut scenarios is typically too broad to provide a good fit to the experimental data \cite{Huber}. Both kind of effects can be adressed in a general framework of effective Lorentz-
and CPT-violation.
\par
In this paper we study the interplay of generic CPT-violating contributions to the neutrino oscillation Hamiltonian, which are capable of mimicking matter effects, and scale linearly with the neutrino energy and altered
dispersion relations in the sterile sector with a different, namely quadratical, energy dependence in accord with
Lorentz-violating and/or extra-dimensional shortcut models for a $2 \times 2$ system of active and sterile neutrino flavors. We find that 
such an approach naturally breaks the CP and CPT invariance of shortcut models and leads to narrower resonance width, which implies
it can be relevant in explaining apparent neutrino oscillation anomalies.
\par
In section \ref{matter} we set the stage for our analysis by reviewing the basic features of CPT-violating
extensions of the standard neutrino oscillation phenomenology and by choosing a set of parameters motivated by the
experimental needs as discussed above.
In section \ref{adr} we show that the different energy dependences for active and sterile flavors 
lead to new resonances in active-sterile neutrino mixing. The resonance
energies in this model depend on the underlying mass hierarchy of active and sterile neutrinos as well as 
on the relative sign of the novel potential terms, i.e. we obtain different resonance energies for neutrinos and
antineutrinos. It is even possible to establish two resonances in the oscillation probability
with a specific mass hierarchy. Moreover, we notice that the interplay between different energy dependences 
can lead to a narrower resonance width as compared to minimal extra-dimensional models. 
\par
In section \ref{discussion} we discuss our results and briefly comment on possible extensions of our model.

\section{A model for active-sterile neutrino oscillations with altered dispersion relations}\label{matter}

In Ref. \cite{Kostelecky:2003cr} a minimal extension of standard model physics
including all possible CPT and Lorentz symmetry-violating Dirac- and Majorana-type couplings to left- and right-handed
neutrinos has been introduced. It provides an effective Hamiltonian $h_{\text{eff}}$ describing neutrino oscillations. An explicit expression for $h_{\text{eff}}$ can be derived by assuming that low-energy physics is dominated by renormalizable
Lorentz-violating operators. In the following considerations we restrict ourselves to the propagation of
one active and one sterile neutrino; we will justify this approach in the upcoming analysis. 
The emergent two-state system is governed by the Schr\"odinger equation
\be
   i\frac{d}{dt} \left(\begin{array}{c} \nu_{\text{a}}(t) \\ \nu_{\text{s}}(t) \end{array}\right) 
   = h_{\text{eff}} \left(\begin{array}{c} \nu_{\text{a}}(t) \\ \nu_{\text{s}}(t) \end{array}\right).
\ee
The index $a$ stands for the active neutrino which can be thought of as an electron, muon or tau neutrino; 
$s$ refers to the sterile neutrino. The effective Hamiltonian in this context can be written as
   \be
      h_{\text{eff}} = \begin{pmatrix}
      \mathfrak{A}_{\text{aa}} & \mathfrak{A}_{\text{as}} \\  \mathfrak{A}_{\text{sa}} & \mathfrak{A}_{\text{ss}}
      \end{pmatrix},
   \ee
where the entries are given by
   \be
      \mathfrak{A}_{\text{ij}} = E \delta_{ij} + \frac{1}{2E} (m_l^{} m_l^{\dagger})_{ij} + \frac{1}{E} \left[
      \left(a_{\text{L}}\right)^{\mu}p_{\mu} - \left(c_{\text{L}}\right)^{\mu\nu}p_{\mu}
      p_{\nu}\right]_{ij} \label{Afrak}
   \ee
with $i,j = a, s$.
$\left(a_{\text{L}}\right)^{\mu}_{ij}$ are CPT- and Lorentz-violating coefficients, whereas $\left(c_{\text{L}}\right)^{\mu\nu}_{ij}$ are CPT-conserving, but Lorentz-violating. $E$ is the
neutrino energy and $p_{\mu}$ its momentum. The usual Lorentz-conserving
mass term in Eq. (\ref{Afrak}) may be parametrized in analogy to two flavor Lorentz-conserving neutrino oscillations
in vacuo
\be
        (m_l^{} m_l^{\dagger})_{ij}^{} = (m_l^{} m_l^{\dagger})_{ij}^{\ast} =
        \frac{\Sigma m^2}{2} + \frac{\Delta m^2}{2} \begin{pmatrix} -\cos2\theta & \sin2\theta \\
        \sin2\theta & \cos2\theta \end{pmatrix},
    \ee
where $\Sigma m^2 = m_1^2 + m_2^2$ and $\Delta m^2 = m_2^2 - m_1^2$. $\Delta m^2$ is the mass splitting
between one of the active neutrinos and the sterile neutrino. The mass splitting between one (or more) sterile state(s) and the three active states is taken to be $\Delta m^2 \approx \Delta m^2_{\text{LSND}} \approx 1 \text{eV}^2$ and thus we have $\Delta m^2 \gg \Delta m^2_{\text{sun}}, ~ \Delta m^2_{\text{atm}}$. In the 3 + 1 neutrino spectrum which is usually discussed as a solution for the 
LSND anomaly only the largest oscillation frequency corresponding to $\Delta m^2_{14} \equiv \Delta m^2$ contributes to neutrino
oscillations on short baselines. Thus a two neutrino scenario is a reasonable approximation. In astrophysical environments also other oscillation frequencies can contribute, but they are unaffected by altered dispersion
relations only affecting the sterile neutrino which dominates the isolated state $\nu_4$. 
Since $p_{\mu}$ in a model with light 
propagating neutrino states can be identified with $p_{\mu} = (E, -\vec{p}) \simeq E (1, -\hat{\vec{p}}) = 
E \hat{p}_{\mu}$, where $\hat{\vec{p}}$ and $\hat{p}_{\mu}$ denote unit vectors, we are now prepared to
explicitly make our choices for specific combinations of non-vanishing $\left(a_{\text{L}}\right)^{\mu}_{ij}$ and
$\left(c_{\text{L}}\right)^{\mu\nu}_{ij}$ coefficients which is mainly guided by the model proposed in Ref. 
\cite{Pas:2005rb} to explain the LSND result~\footnote{In models with Lorentz-violating interactions the notion of a preferred direction emerges and it is convenient to choose a specific coordinate system to report experimental findings. Here we choose a Sun-centered celestial equatorial frame with coordinates $(T, X, Y, Z)$. In this frame the Earth's rotational axis coincides with the $Z$ direction and a parametrization $ \hat{\vec{p}} = \left(\sin\Theta \cos\Phi,~\sin\Theta \sin\Phi,~\cos\Theta\right)$,
where $\hat{\vec{p}}$ is the unit 3-vector of the particle's momentum and $\Theta$ is the celestial
colatitude and $\Phi$ denotes the celestial longitude is convenient.}. As has been noticed in the aforementioned reference
the MiniBooNE neutrino oscillation signal could be explained by an extra-dimensional model with neutrino 
oscillations involving active and sterile states (details of such models will be explained later). 
The recent results of the MiniBooNE collaboration for the
antineutrino channel, however, challenge all explanations relying on gravitationally-driven non-standard neutrino
oscillations. Those models are by construction CP-symmetric and thus cannot accommodate experimental 
CP-antisymmetric observations. It is therefore promising to keep the relevant features of shortcut models, 
such as the resonant neutrino oscillations, and extend the idea by incorporating physics which is known to
be CP-antisymmetric. The latter is, as one would naively conjecture, most conveniently achieved by looking at 
matter effects or, as we shall henceforth do, generic CPT-violating effects which mimic matter effects. 
\par\noindent
\newline
In a first step we are concerned with specifying the $a_{\text{L}}$-type coefficient in a phenomenologically
relevant way.
We assume that only $\left(a_{\text{L}}\right)^{\mu}_{aa} \neq 0$ and we may therefore write
   \be
      \frac{1}{E} \left(a_{\text{L}}\right)^{\mu}_{ij}p_{\mu} = \frac{1}{2E} 2
      \left(a_{\text{L}}\right)^{\mu}_{aa}\hat{p}_{\mu}E.
   \ee
Moreover, we introduce two new variables $a$ and $A$ in order to streamline notation and for practical purposes when it comes to analyzing the emergent new resonance structures in active-sterile neutrino oscillations:
   \be
      A(E) = aE, \qquad \qquad a = 2\left(a_{\text{L}}\right)^{\mu}_{aa}\hat{p}_{\mu} \label{Aanda}.
   \ee
At this stage of our analysis a few comments are in order. Our choice for $\left(a_{\text{L}}\right)^{\mu}_{aa}$
to be the only non-vanishing $a_{\text{L}}$-type coefficient is guided by the fact that it 
contains the $V-A$ coupling for coherent elastic forward scattering of neutrinos in matter 
\cite{Wolfenstein:1977ue, Mikheev:1986wj} encoded in
$\left(a_{\text{L}}\right)^{0}_{aa}$. If the active neutrino is propagating through matter this coefficient 
could therefore be interpreted as being proportional to the electron $n_e$, neutron $n_n$ and proton $n_p$ number densities
which cause forward neutral and charged current exchange scattering processes for the neutrinos. 
Assuming a with respect to the electric charge ($n_e \approx n_p$) neutral matter accumulation we can identify
   \be
      \left(a_{\text{L}}\right)^{0}_{ee} &=& \sqrt{2}G_{\text{F}}\left(n_e - 
      \frac{1}{2}n_n\right), \label{a0e}\\
      \left(a_{\text{L}}\right)^{0}_{\mu\mu} = \left(a_{\text{L}}\right)^{0}_{\tau\tau}
      &=& \sqrt{2}G_{\text{F}}\left(-\frac{1}{2}n_n\right) \label{a0m}.
   \ee
Since we are primarily concerned with the phenomenology of earth-bound neutrino oscillation experiments 
it is convenient to assume matter composed of light elements, i.e. $n_p \approx n_n$, which further
simplifies Eqs. (\ref{a0e}, \ref{a0m}). It is obvious that in this case the matter potential $A$ discriminates
between flavors in a way that opposite signs are obtained for electron neutrinos on the one hand and muon and
tau neutrinos on the other. 
\par
Moreover, it is by no means necessary to interpret the coefficient $\left(a_{\text{L}}\right)^{0}_{aa}$
as a matter potential. It is always possible to regard it as a generic CPT- and Lorentz-violating coefficient.
This coefficient can then either have the same sign as a possible matter potential term (flavor-aligned case)
or the opposite sign (flavor-misaligned case) if it discriminates between flavors. 
It is also possible that it does not discriminate between flavors at all (flavor-blind case).
\par
In principle the variable $a$ is also direction dependent. This dependence is introduced via the occurrence
of the neutrino's 3-momentum in Eq. (\ref{Aanda}b). However, the only thing which matters for the resonance structures is the energy dependence of the variable $A$. 
Since the contraction of the Lorentz indices in $a$ involves a unit vector it does not entail another
energy dependence, i.e. in this parametrization $A$ does scale linearly with the neutrino energy $E$.
\par
So our model can accommodate matter effects and generic CPT violation
at the same time, i.e. it might be relevant for both long and short baseline neutrino oscillation experiments. 
\par\noindent
\newline
In a second step we now comment on the role of the $c_{\text{L}}$-type coefficients. Here
we choose the sterile neutrino coefficient $\left(c_{\text{L}}\right)^{\mu\nu}_{ss}$ to be the only non-vanishing component. We thus write
   \be
      \frac{1}{E} \left(c_{\text{L}}\right)^{\mu\nu}_{ij}p_{\mu}p_{\nu} = 
      \frac{1}{2E} 2\left(c_{\text{L}}\right)^{\mu\nu}_{ss}\hat{p}_{\mu}\hat{p}_{\nu} E^2
   \ee
and introduce two new variables
   \be
      B(E) = b E^2, \qquad b = 2\left(c_{\text{L}}\right)^{\mu\nu}_{ss}\hat{p}_{\mu}\hat{p}_{\nu}.
   \ee
In this parametrization the coefficient $B$ scales quadratically with the neutrino energy and we will motivate
shortly how this energy dependence arises naturally also in models with sterile neutrinos taking shortcuts
in extra spatial dimensions. Before we delve into this subject, we note that we can impose further restrictions
on $\left(c_{\text{L}}\right)^{\mu\nu}$. The trace $\eta_{\mu\nu}\left(c_{\text{L}}\right)^{\mu\nu}_{ij}$
is unobservable due to the fact that it may be absorbed into the usual kinetic term of the Hamiltonian, it
is thus irrelevant in neutrino oscillation experiments and may be set to zero. Note, moreover, that the 
$c_{\text{L}}$-type coefficients are also direction dependent and $b$ can have either sign if $\left(c_{\text{L}}\right)^{\mu\nu}_{ij}$
is thought of as a generic CPT-conserving but Lorentz-violating coefficient.
Again we neglect the direction dependence for the moment since we are primarily interested in the 
resonance phenomenology.
\par
Let us now review the key features of active-sterile neutrino oscillations in models with extra spatial
dimensions. We focus on a $3+1$ dimensional Minkowskian brane, to which the standard model 
particles are confined, embedded in an extra-dimensional bulk \cite{ArkaniHamed:1998rs, Randall:1999ee}. 
Singlets under the gauge group, like 
sterile neutrinos, may travel freely on the brane as well as in the bulk. Consider for example a
sterile neutrino which propagates in the bulk as well as on the brane. We are assuming in our model 
that the sterile neutrinos behave like classical point particles. It is then possible to
calculate the travel time of the neutrino on the brane and in the bulk via the geodesic equations.
The geodesic equations in turn explicitly contain the metric of the underlying spacetime encapsulated in
the Christoffel connection coefficients and in that sense extra-dimensional shortcut physics is model-dependent.
It has been shown that for certain spacetimes, like a spacetime exhibiting small brane fluctuations \cite{Pas:2005rb} or an asymmetrically-warped spacetime \cite{Hollenberg:2009ws}, the travel time in the bulk $t^{\text{bulk}}$ will be smaller than the travel time on the brane $t^{\text{brane}}$. Moreover, since an active neutrino flavor is confined to the brane, it appears that the sterile neutrino may take a shortcut through an extra dimension as seen from the brane. The difference in travel times is encoded in the dimensionless shortcut parameter $\epsilon$, 
which is defined as
    \be
        \epsilon = \frac{t^{\text{brane}}-t^{\text{bulk}}}{t^{\text{brane}}}.
    \ee
The shortcut through an extra dimension can be parametrized via the effective potential $b = 2\epsilon$. 
Note that this geometric model relies on metric shortcuts and that the equations of motion are identical for particles and antiparticles. Thus the model prediction are CP-symmetric but Lorentz-violating.
In addition to that shortcut models single out one specific choice for the sign of $b$, namley $b > 0$,
and thus represent a special class of Lorentz-violating extensions of the standard neutrino oscillation
picture. 
\par\noindent
For a detailed analysis of shortcut physics and the shortcut parameter $\epsilon$ the reader is referred to Refs. \cite{Pas:2005rb, Hollenberg:2009ws}.
\par\noindent
\newline
We are now prepared to write down the effective Hamiltonian for our model:
   \be
      h_{\text{eff}} = \frac{\Sigma m^2}{4E} + \frac{\Delta m^2}{4E} \begin{pmatrix} -\cos2\theta & \sin2\theta \\
        \sin2\theta & \cos2\theta \end{pmatrix} + \frac{1}{2E} \begin{pmatrix} A(E)
         & 0 \\ 0 & -B(E) \end{pmatrix} \label{heff}.
   \ee
Since the forthcoming analysis deals with resonant neutrino oscillations it is important to know the 
energy dependence of the effective potentials $A$ and $B$. An explicit choice of coefficients $\left(a_{\text{L}}\right)^{\mu}_{aa}$ and $\left(c_{\text{L}}\right)^{\mu\nu}_{ss}$
which goes beyond the assumptions made above is not necessarily called for unless a detailed account
of the direction dependence of the resonance structures is sought. For the purpose of the present paper
a detailed analysis of how direction dependence effects influence, i.e. suppress or enhance, the resonant 
structures seems unwarranted since it strongly depends on the explicit choice of non-vanishing components
in $\left(a_{\text{L}}\right)^{\mu}_{aa}$ and $\left(c_{\text{L}}\right)^{\mu\nu}_{ss}$. We note, however, that
besides the resonant enhancement of events in oscillation experiments triggered by resonances in the
oscillation probability the breakdown of rotational symmetry also yields interesting and testable sidereal
variations \cite{Diaz:2009qk,Barger:2007dc}. Such variations, up to now, have not been detected, 
whereas the reported excesses of events in LSND and MiniBooNE might indicate resonant oscillations.

\section{Resonant active-sterile neutrino oscillations}\label{adr}

Given the effective Hamiltonian $h_{\text{eff}}$ of Eq. (\ref{heff}) it is straightforward to diagonalize this
two-state system by introducing effective mass eigenstates
   \be
      \left(\begin{array}{c} \nu_{\text{a}} \\ \nu_{\text{s}}\end{array}\right) = \begin{pmatrix}
        \cos\tilde\theta & \sin\tilde\theta \\
        -\sin\tilde\theta & \cos\tilde\theta \end{pmatrix} \left(\begin{array}{c} \tilde\nu_1
        \\ \tilde\nu_2 \end{array}\right), \label{PMNS}
   \ee
where the diagonal Hamiltonian is obtained via
   \be
      h_{\text{eff}}^{\text{diag}} = R^{\dagger} h_{\text{eff}} R.
   \ee
$R$ is the rotation matrix introduced in Eq. (\ref{PMNS}); $\tilde\theta$ is an effective mixing angle,
$\tilde\nu_1$ and $\tilde\nu_2$ are effective mass eigentates in the presence of $A$ and $B$ potentials.
The effective mixing angle is calculated to obey
   \be
      \tan2\tilde\theta = \frac{\tan2\theta}{1-\frac{A+B}{\Delta m^2 \cos2\theta}} \label{tantilde}.
   \ee
One readily infers from Eq. (\ref{tantilde}) that for combined $A$- and $B$-type potentials multiple
resonant active-sterile neutrino mixing solutions become possible. The condition for resonant neutrino mixing 
is found to be
   \be
      bE^2 + aE - \Delta m^2 \cos2\theta = 0 \label{resendef}
   \ee
and it is easily solved to give
   \be
      E_{\text{res}} = -\frac{a}{2b} \pm \sqrt{\left(\frac{a}{2b}\right)^2 + 
        \frac{\Delta m^2 \cos2\theta}{b}} \label{resen}.
   \ee
Resonant neutrino mixing yields three distinct energy domains: Below the resonance energy vacuum mixing is recovered,
at the resonance energy the effective mixing becomes maximal ($\tilde\theta = \pi/4$), while above the resonance
active and sterile neutrinos decouple and oscillations are suppressed.
The relation for the resonance energy must now be exploited in the different scenarios of flavor-alignment, flavor-misalignment and flavor-blindness. By inspecting Eq. (\ref{resen}) we already notice at this stage
of our analysis that a large variety of different neutrino oscillation phenomenologies (different flavors, 
different signs for $a$ and $b$, different mass hierarchies between one of the three active and the sterile
neutrino) can be subsumed in four distinct possible resonance phenomenologies. 
\par
However, before we delve into the details here it is instructive to rederive the well-understood resonances
for the MSW \cite{Wolfenstein:1977ue, Mikheev:1986wj} and extra-dimensional shortcut scenarios \cite{Pas:2005rb, Hollenberg:2009ws}.
The former is readily obtained by setting $b = 0$ in Eq. (\ref{resendef}) and invoking 
Eqs. (\ref{a0e}, \ref{a0m}). This yields
   \be
      E_{\text{res}} = \frac{\Delta m^2 \cos2\theta}{a} = \frac{\Delta m^2 \cos2\theta}{\pm\sqrt{2}G_{\text{F}}n_e}
      \label{exmat},
   \ee
where for the last equals sign an electron flavor is assumed for illustrative purposes. This
choice does indeed reflect the usual MSW resonance except for we are dealing with one active and one sterile 
flavor. In this case of a matter resonance for an electron flavor Eq. (\ref{exmat}) reveals resonant neutrino 
mixing for two different cases. On the one hand, for neutrinos, the sign in the denominator is positive and we therefore need $\text{sgn}~\Delta m^2 = \text{sgn}~\cos2\theta$ to obtain resonant mixing. On the other hand
it is possible to generate resonant antineutrino mixing in matter, if the mass hierarchy is inverted, i.e.
$\text{sgn}~\Delta m^2 \neq \text{sgn}~\cos2\theta$. Note, however, that it is not possible to obtain resonant 
mixing for both neutrinos and antineutrinos at the same time; matter effects break CP invariance. For muon and tau neutrinos the situation is reversed.
\par
In the framework of active-sterile neutrino oscillations with extra-dimensional shortcuts only we readily get
   \be
      E_{\text{res}} = \sqrt{\frac{\Delta m^2 \cos2\theta}{b}}.
   \ee
In this context $b$ is a positive quantity and we need a mass hierarchy $\text{sgn}~\Delta m^2 = \text{sgn}~\cos2\theta$ for resonant mixing. Here the resonance exists for both neutrinos and antineutrinos
and for all flavors at the same time which resembles the CP symmetry of such models discussed above.
\par\noindent
\newline
Let us now turn to Eq. (\ref{resen}) in order to understand the emergent new resonances in detail. It is convenient
to introduce two new variables 
   \be
      \alpha = \frac{a}{2b}, \qquad \beta = \frac{\Delta m^2 \cos2\theta}{b}
   \ee
and rewrite
   \be
      E_{\text{res}} = -\alpha \pm \sqrt{\alpha^2 + \beta}.
   \ee
The relevant information now is whether $\alpha$ and $\beta$ do have positive or negative sign. Four cases are
distinguishable and we find the possible resonance energies:
   \begin{enumerate}
    \item $\text{sgn}~\alpha = +1$, $\text{sgn}~\beta = +1$: \\
          \be
             E_{\text{res}}^{++} = -\left|\alpha\right| + \sqrt{\alpha^2 + \left|\beta\right|}
             \label{resen1}
          \ee
    \item $\text{sgn}~\alpha = -1$, $\text{sgn}~\beta = -1$: \\
          \be
             E_{\text{res}}^{--} = +\left|\alpha\right| \pm \sqrt{\alpha^2 - \left|\beta\right|}
          \label{resen2}
          \ee
          Here the additional condition $\alpha^2 \geq \left|\beta\right|$ has to be satisfied.
    \item $\text{sgn}~\alpha = +1$, $\text{sgn}~\beta = -1$: \\
          \begin{center} 
           In this case no resonance exists.
          \end{center}
          \label{resen3}
    \item $\text{sgn}~\alpha = -1$, $\text{sgn}~\beta = +1$: \\
          \be
             E_{\text{res}}^{-+} = +\left|\alpha\right| + \sqrt{\alpha^2 + \left|\beta\right|}
          \label{resen4}
          \ee
   \end{enumerate}
Note that $E_{\text{res}}^{-+} > E_{\text{res}}^{++}$. This will be relevant for possible neutrino
oscillation phenomenologies to be discussed later.
\begin{figure}
\centering 
%\raisebox{6cm}{$m_{\text{eff}}^2[\text{eV}^2]$} 
\includegraphics[scale=1.2]{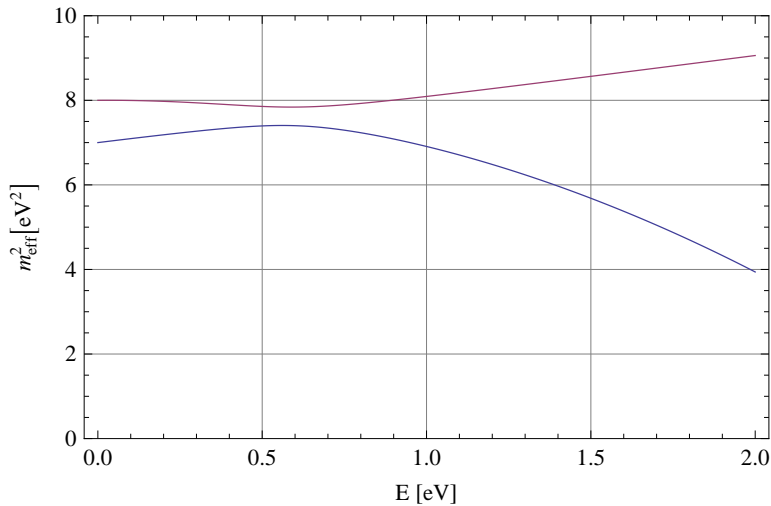}
%\\{\hspace{10cm}$E[\text{eV}]$}
\caption{The effective mass squared as a function of the energy for the case of $E^{++}_{\text{res}}$. We choose
$a=1~\text{eV}$, $b=1$, $\cos2\theta=0.9$, $\Delta m^2=1~\text{eV}^2$, $\Sigma m^2=15~\text{eV}^2$ for illustrative purposes. The red curve (top) corresponds to $m_{2, \text{eff}}^2$, the blue curve (bottom) corresponds to $m_{1, \text{eff}}^2$. The minimal distance between the two curves corresponds to the position of the resonance.} 
\label{a+1m+1}
\end{figure}

%\begin{figure}
%\centering 
%%\raisebox{6cm}{$m_{\text{eff}}^2[\text{eV}^2]$} 
%\includegraphics[scale=1.2]{Fig2.eps}
%%\\{\hspace{10cm}$E[\text{eV}]$}
%\caption{The effective mass squared as a function of the energy for the case (1b). We chose
%$a=-1~\text{eV}$, $b=1$, $\cos2\theta=0.9$, $\Delta m^2=1~\text{eV}^2$, $\Sigma m^2=15~\text{eV}^2$ for illustrative %purposes. The red curve (top) corresponds to $m_{2, \text{eff}}^2$, the blue curve (bottom) corresponds to $m_{1, %\text{eff}}^2$.
%The minimal distance between the two curves corresponds to the position of the resonance.} 
%\label{a-1m+1}
%\end{figure}                        
            
\begin{figure}
\centering 
%\raisebox{6cm}{$m_{\text{eff}}^2[\text{eV}^2]$} 
\includegraphics[scale=1.2]{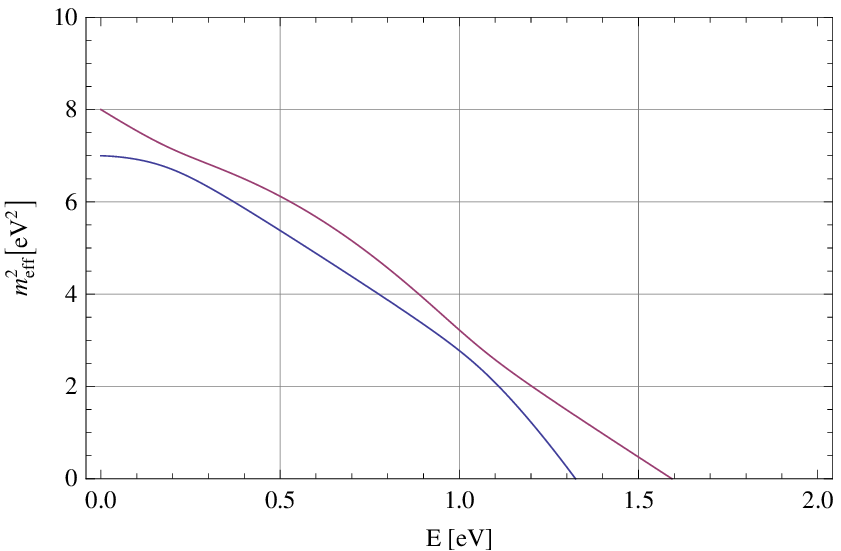}
%\\{\hspace{10cm}$E[\text{eV}]$}
\caption{The effective mass squared as a function of the energy for the case $E^{--}_{\text{res}}$ with 
$\alpha^2 > \left|\beta\right|$. We chose
$a=-5~\text{eV}$, $b=4$, $\cos2\theta=0.9$, $\Delta m^2=-1~\text{eV}^2$, $\Sigma m^2=15~\text{eV}^2$ for illustrative purposes. The blue curve (bottom) corresponds to $m_{2, \text{eff}}^2$, the red curve (top) corresponds 
to $m_{1, \text{eff}}^2$.
The minimal distances between the two curves correspond to the positions of the resonances. Note that the resonance
condition is realized for two different values of the neutrino energy.}
\label{a-1m-1twores}
\end{figure}
                     
%\begin{figure}
%\centering 
%%\raisebox{6cm}{$m_{\text{eff}}^2[\text{eV}^2]$} 
%\includegraphics[scale=1.2]{Fig4.eps}
%%\\{\hspace{10cm}$E[\text{eV}]$}
%\caption{The effective mass squared as a function of the energy for the case (2b) with $
%\left(\frac{|a|}{2b}\right)^2 = \frac{|\Delta m^2 \cos2\theta|}{b}$. We chose
%$a=-4~\text{eV}$, $b=4$, $\cos2\theta=0.9$, $\Delta m^2=-1~\text{eV}^2$, $\Sigma m^2=15~\text{eV}^2$ for %illustrative purposes. The blue curve (bottom) corresponds to $m_{2, \text{eff}}^2$, the red curve (top) corresponds %to $m_{1, \text{eff}}^2$.
%The minimal distance between the two curves corresponds to the position of the resonance.} 
%\label{a-1m-1}
%\end{figure}
The underlying physics of the new resonances can be understood intuitively as well. We illustrate this for
the case of a scenario with extra-dimensional shortcuts: Due to the fact
that $B$ is positive here and the term describing shortcut-like effects has negative sign, the
mass eigenstates of sterile neutrinos in the presence of shortcut potentials are lowered with 
increasing energy, regardless of the particle or antiparticle nature of the neutrinos. 
For the case of neutrinos in which $\text{sgn}~a = +1$ (electron neutrinos in the case of flavor-alignment or
muon/tau neutrinos for flavor-misalignment), the mass eigenstates of the active flavor 
are raised and thus a resonance occurs for a mass hierarchy where $\text{sgn}~\Delta m^2 =  +1$. 
For the inverted mass hierarchy with $\text{sgn}~\Delta m^2 =  -1$ the mass eigenstates are consequently 
moving away from each other for increasing neutrino energy and a resonance is no longer possible.

\subsection{Flavor-alignment}

We now turn to the question which resonance structures one encounters for the different cases of
flavor-alignment, flavor-misalignment and flavor-blindness. We start by reviewing the flavor-aligned case
in which the $a_{\text{L}}$-type coefficient can be thought of as or at least mimicking the common matter 
effects. For reasons of bookkeeping we introduce a notation in which we put all relevant information as
   \be
      \left(\text{neutrino flavor}, \text{sgn}~a, \text{sgn}~b, \text{sgn}~\Delta m^2\right).
   \ee
All possible combinations for electron neutrinos in the flavor-aligned case are then given by
   \be
      &&\left(\nu_{e}, +, +, + \right) \qquad \left(\bar{\nu}_{e}, -, +, +\right) \nonumber \\
      &&\left(\nu_{e}, +, +, - \right) \qquad \left(\bar{\nu}_{e}, -, +, -\right) \nonumber \\
      &&\left(\nu_{e}, +, -, + \right) \qquad \left(\bar{\nu}_{e}, -, -, +\right) \nonumber \\
      &&\left(\nu_{e}, +, -, - \right) \qquad \left(\bar{\nu}_{e}, -, -, -\right) \nonumber 
   \ee
whereas for muon and tau neutrinos we have
   \be
      &&\left(\nu_{\mu,\tau}, -, +, + \right) \qquad \left(\bar{\nu}_{\mu,\tau}, +, +, +\right) \nonumber \\
      &&\left(\nu_{\mu,\tau}, -, +, - \right) \qquad \left(\bar{\nu}_{\mu,\tau}, +, +, -\right) \nonumber \\
      &&\left(\nu_{\mu,\tau}, -, -, + \right) \qquad \left(\bar{\nu}_{\mu,\tau}, +, -, +\right) \nonumber \\
      &&\left(\nu_{\mu,\tau}, -, -, - \right) \qquad \left(\bar{\nu}_{\mu,\tau}, +, -, -\right) \nonumber.
   \ee
As has been mentioned earlier these combinations still carry some redundancies and can be reduced further. We do
this by introducing another shorthand notation
   \be
      \left(\text{neutrino flavor}, \text{sgn}~\alpha, \text{sgn}~\beta\right) \label{shortres}.
   \ee
The possible resonance structures for the different cases then become more transparent since
in Eqs. (\ref{resen1}-\ref{resen4}) we introduced a notation $E_{\text{res}}^{\text{sgn}~\alpha ~ \text{sgn}~\beta}$ 
for the emergent resonant energies. The notation (\ref{shortres}) therefore contains information on the
resonance energies for the different flavors as well as particles and antiparticles.
For electron neutrinos we find
   \be
      &&\left(\nu_{e}, +, +, + \right) ~ \to ~ \left(\nu_{e}, +, + \right) \qquad \left(\bar{\nu}_{e}, -, +, +\right) ~ \to ~ \left(\bar{\nu}_{e}, -, + \right) \nonumber \\
      &&\left(\nu_{e}, +, +, - \right) ~ \to ~ \left(\nu_{e}, +, - \right) \qquad \left(\bar{\nu}_{e}, -, +, -\right) ~ \to ~ \left(\bar{\nu}_{e}, -, - \right) \nonumber \\
      &&\left(\nu_{e}, +, -, + \right) ~ \to ~ \left(\nu_{e}, -, - \right) \qquad \left(\bar{\nu}_{e}, -, -, +\right) ~ \to ~ \left(\bar{\nu}_{e}, +, - \right) \nonumber \\
      &&\left(\nu_{e}, +, -, - \right) ~ \to ~ \left(\nu_{e}, -, + \right) \qquad \left(\bar{\nu}_{e}, -, -, -\right) ~ \to ~ \left(\bar{\nu}_{e}, +, + \right) \nonumber.
   \ee
For muon and tau neutrinos this yields
   \be
      &&\left(\nu_{\mu,\tau}, -, +, + \right) ~ \to ~ \left(\nu_{\mu,\tau}, -, + \right) \qquad \left(\bar{\nu}_{\mu,\tau}, +, +, +\right) ~ \to ~ \left(\bar{\nu}_{\mu,\tau}, +, + \right) \nonumber \\
      &&\left(\nu_{\mu,\tau}, -, +, - \right) ~ \to ~ \left(\nu_{\mu,\tau}, -, - \right) \qquad \left(\bar{\nu}_{\mu,\tau}, +, +, -\right) ~ \to ~ \left(\bar{\nu}_{\mu,\tau}, +, - \right) \nonumber \\
      &&\left(\nu_{\mu,\tau}, -, -, + \right) ~ \to ~ \left(\nu_{\mu,\tau}, +, - \right) \qquad \left(\bar{\nu}_{\mu,\tau}, +, -, +\right) ~ \to ~ \left(\bar{\nu}_{\mu,\tau}, -, - \right) \nonumber \\
      &&\left(\nu_{\mu,\tau}, -, -, - \right) ~ \to ~ \left(\nu_{\mu,\tau}, +, + \right) \qquad \left(\bar{\nu}_{\mu,\tau}, +, -, -\right) ~ \to ~ \left(\bar{\nu}_{\mu,\tau}, -, + \right) \nonumber.
   \ee
We can now again subdivide our results into three different cases: For a given (anti-)particle of a fixed flavor
there are either none, one or two resonances possible. Let us illustrate this point for the electron
neutrino: In the case of $\left(\nu_{e}, +, +, + \right)$  
we find one resonance energy given by 
$E_{\text{res}}^{++}$; for $\left(\nu_{e}, +, +, - \right)$ no resonance exists; and finally for 
$\left(\nu_{e}, +, -, - \right)$ there might even be two resonances at the same time with energies 
$E_{\text{res}}^{--}$, if the condition $\alpha^2 \geq \left|\beta\right|$ holds. Such is the situation for electron 
neutrinos only. In the light of the reported LSND and MiniBooNE neutrino oscillation anomalies the
behavior of the resonance structures under CP conjugation is of special interest. As has been mentioned the
MiniBooNE collaboration reports an anomalous excess of events in the neutrino channel, but does not find any
deviation from the standard oscillation picture for the antineutrino data. This excess of events can be generated
by resonant neutrino oscillations as has been shown earlier (see, e.g., Ref. \cite{Pas:2005rb}) and for this reason we are foremost interested in setups which allow for different resonance structures for neutrinos and
antineutrinos. In our example of electron neutrinos this can indeed be established. By choosing a parameter 
set $\text{sgn}~b = -1, ~\text{sgn}~\Delta m^2 = +1$ we find that resonant oscillations 
are encountered for electron neutrinos but not for antielectron neutrinos. On the other hand it is also possible 
to obtain resonant mixing for antielectron neutrinos but not for electron neutrinos via $\text{sgn}~b = +1, ~\text{sgn}~\Delta m^2 = -1$. The novel resonance structures encountered in our model are thus capable of matching
the observed excess of events in the LSND and/or MiniBooNE oscillation experiments at least qualitatively since the resonant behavior, encoded in $\sin2\tilde\theta$, persists in the oscillation probability as written in Eq. (\ref{Posc}).
The intriguing feature of such CPT-violating approaches is that they might be relevant for earth-bound neutrino oscillation experiments.
Eventually, we note that the resonance structures for electron neutrinos coincide with those of antimuon and -tau
neutrinos as well as the resonances for antielectron neutrinos are the same as those for muon and tau neutrinos.
Note, however, that $a$ is an abbreviatory notation and does contain an additional flavor dependence via 
$\left(a_{\text{L}}\right)^{\mu}_{aa}$. Thus one would not necessarily expect the resonance energies for the
different flavors to be the same which again enriches the possible implications of our model and on the
other hand reveals the necessity for a detailed study of the world's neutrino oscillation data in our framework.
\par
As a concluding remark we state that it is by no means necessary for the muon and tau neutrinos to have the same sign
when it comes to the $a_{\text{L}}$-type coefficient. All other combinations are equally possible, i.e. 
$\text{sgn}~\left(a_{\text{L}}\right)^{\mu}_{ee} = \text{sgn}~\left(a_{\text{L}}\right)^{\mu}_{\mu\mu} \neq \text{sgn}~\left(a_{\text{L}}\right)^{\mu}_{\tau\tau}$ for example.
\par
The energy dependences of the effective mass-squared differences are depicted in
Figs. \ref{a+1m+1} and \ref{a-1m-1twores}. The resonances are encountered for the 
energy for which the effective mass gap between the two curves is minimal.

\subsection{Flavor-misalignment}

Flavor-misalignment describes the case in which the sign for the $a_{\text{L}}$-type coefficient is reversed with
respect to the flavor-aligned case. Thus the sign of $a$ does not mimic common matter effects any more, but the
opposite situation when it comes to the flavor structure. We find that the resonance structures calculated for electron, muon and tau neutrinos now hold for antineutrinos and vice versa. There is another interesting consequence for potential signals in neutrino oscillation experiments. For the case of flavor-alignment we found that the resonance energy for electron 
neutrinos, $E^{++}_{\text{res}}$, is smaller than the resonance energy for antielectron neutrinos, 
$E^{-+}_{\text{res}}$. This situation is now also reversed. One would expect a resonance energy 
$E^{-+}_{\text{res}}$ for antielectron neutrinos and $E^{++}_{\text{res}}$ for electron neutrinos depending on the 
sign of $\alpha$. The difference
in resonance energies for both cases is calculated to be 
   \be
      \left|\Delta E^{\nu\bar{\nu}}\right| = \left| E^{-+}_{\text{res}} - E^{++}_{\text{res}} \right|
      = 2 \left|\alpha\right|.
   \ee
The prediction of a resonance in both neutrino and antineutrino oscillations thus allows to extract information
on the magnitude of the ratio of the CPT-violating coefficients.
\par
Note, however, that there is still some residual ambiguity which stems from the fact that
depending on the sign of $\alpha$ for both the flavor-aligned as well as the flavor-misaligned case one either finds 
the electron or antielectron resonance energy to be larger. One can therefore not easily distinguish
between flavor-aligned and flavor-misaligned physics.
\par
We will give an order of magnitude estimate for the CPT-violating coefficients in section \ref{discussion} for
different physics cases.

\subsection{Flavor-blindness}

Flavor-blindness gives the case in which the sign for the potential $a$ has the same sign for all flavors.
Since the aforementioned potential also includes another flavor-dependence via $\left(a_{\text{L}}\right)^{\mu}_{aa}$
the emergent term in the Hamiltonian is not just flavor-diagonal and can thus not simply be ignored when it
comes to analyzing its impact on neutrino oscillations. In particular we notice that the resonances for the different flavors are thus not necessarily identical. For the case of neutrinos we find
   \be
      &&\left(\nu_{e, \mu, \tau}, \pm, +, + \right) ~ \to ~ \left(\nu_{e, \mu, \tau}, \pm, + \right)  \nonumber \\
      &&\left(\nu_{e, \mu, \tau}, \pm, +, - \right) ~ \to ~ \left(\nu_{e, \mu, \tau}, \pm, - \right)  \nonumber \\
      &&\left(\nu_{e, \mu, \tau}, \pm, -, + \right) ~ \to ~ \left(\nu_{e, \mu, \tau}, \mp, - \right)  \nonumber \\
      &&\left(\nu_{e, \mu, \tau}, \pm, -, - \right) ~ \to ~ \left(\nu_{e, \mu, \tau}, \mp, + \right)  \nonumber,
   \ee
whereas for antineutrinos the sign of $a$ has to be reversed and $\text{sgn}~\alpha$, $\text{sgn}~\beta$ have to
be changed accordingly.

\subsection{Resonance widths}

The new eigenvalue difference for the Hamiltonian $h_{\text{eff}}$ is calculated to be
    \be
        \delta h \equiv \frac{\Delta m^2_{\text{eff}}}{2E} = 
        \frac{\Delta m^2}{2E} \sqrt{\left(\frac{A}{\Delta m^2}-\cos2\theta\right)^2 +
        \left(\frac{B}{\Delta m^2}-\cos2\theta\right)^2 + \frac{2AB}{\left(\Delta m^2\right)^2} +
        1 - 2\cos^22\theta} \label{DHeff}\ , 
    \ee
where it is easily seen that in the limiting case $A \to 0$ or $B \to 0$ we obtain the standard MSW or
shortcut scenario result for the eigenvalue difference of the effective Hamiltonian. In the model under
consideration, $\delta h$ differs from the standard formalism by the square-root factor in Eq. (\ref{DHeff})
and it might thus allow to fit data with a larger $\Delta m^2$ if the resonance energy occurs in the range
of LSND/KARMEN.
It is now straightforward to calculate the oscillation probability of active-sterile neutrino mixing
in the presence of effects induced via $a_{\text{L}}$- and $c_{\text{L}}$-type coefficients which is given by
    \be
        P(\nu_a \to \nu_s) = \sin^22\tilde\theta ~ \sin^2\frac{L\delta h}{2} \label{Posc},
    \ee
where $L$ is the length of the experimental baseline and
    \be
        \sin^22\tilde\theta = \frac{\sin^22\theta}{\left(\frac{A}{\Delta m^2}-\cos2\theta\right)^2 +
        \left(\frac{B}{\Delta m^2}-\cos2\theta\right)^2 + \frac{2AB}{\left(\Delta m^2\right)^2} +
        1 - 2\cos^22\theta} \label{sineff}. 
    \ee
In order to examine the resonance structure of the oscillation probability it is convenient to introduce
the ratio of sines of the effective and standard mixing angle in terms of their energy dependence 
{\small \be
        R(E) &=& \frac{\sin^22\tilde\theta}{\sin^22\theta} \\ 
             &=& \frac{\left(\Delta m^2\right)^2}{\left(aE - \Delta m^2
                 \cos2\theta\right)^2 + 
                 \left(bE^2 - \Delta m^2 \cos2\theta\right)^2 + 2abE^3 + \left(\Delta m^2\right)^2 
                 \left(1 - 2\cos^22\theta\right)}.
        \ee} 
The extremum points of this expression correspond to the resonance energies given by Eq. (\ref{resen}).
\begin{figure}
\centering 
%\raisebox{6cm}{$\sin^22\theta_{\text{ms}}$} 
\includegraphics[scale=1.2]{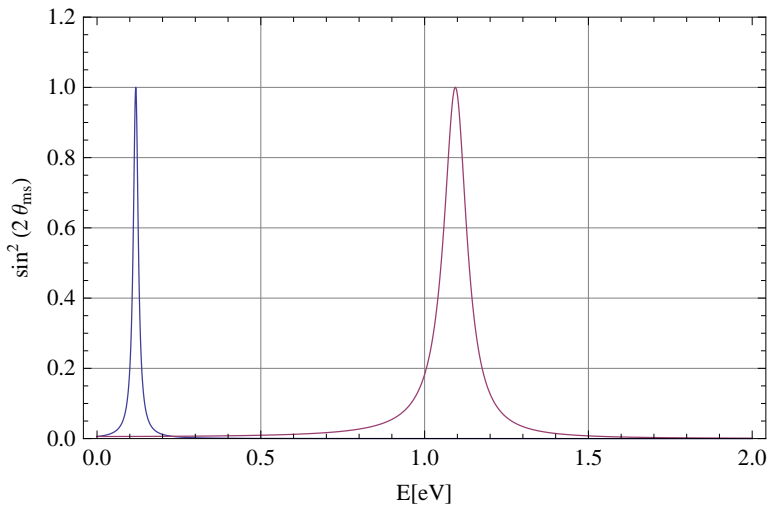}
%\\{\hspace{10cm}$E[\text{eV}]$}
\caption{Energy dependence of the sine of the effective mixing angle $\tilde\theta$ for neutrinos with
a mass hierarchy $\text{sgn}~\Delta m^2 = +1$. We choose $a=10~\text{eV}$, $b=1$, $\cos^22\theta=0.997$, 
$\Delta m^2=1.2~\text{eV}^2$ for illustrative purposes. The blue curve (left) corresponds to the case of combined 
$a_{\text{L}}$- and $c_{\text{L}}$-type coefficients; the red curve (right) shows the resonance for $c_{\text{L}}$ effects only. It is obvious that the resonance width for combined $A$ and $B$ potentials is narrower as compared to $B$ potentials only. Moreover the resonance for combined potentials is shifted to lower energies as both potentials contribute.} 
\label{sina+1m+1}
\end{figure}
\par
We now turn to the question in which way the presence of $a_{\text{L}}$-type coefficientes modifies 
the width of the resonance for shortcut-like effects in vacuo. This is interesting because shortcut-like
models \cite{Pas:2005rb, Hollenberg:2009ws} have been proposed as one possible solution to the yet unexplained neutrino oscillation anomalies 
such as LSND and MiniBooNE. There have, however, been analyses suggesting
\cite{Huber} that the resonance width found in scenarios with sterile neutrinos taking shortcuts through extra dimensions is typically too broad. This problem does not persist if $a_{\text{L}}$-type
coefficients are included:
\par
It is illustrative to consider the difference in energies at Full Width Half 
Maximum $\Delta E(\text{FWHM})$ for $R(E)$.
We find equal widths for both signs of $a$,
$\text{sgn}~a = \pm 1$, in the presence of $a_{\text{L}}$- and $c_{\text{L}}$-type coefficients
    \be
        \left|\Delta E^{\text{ac}}(\text{FWHM})\right| = \left|\sqrt{\alpha^2 + 
        \beta\left(1 + \tan2\theta\right)}\right. - 
        \left.\sqrt{\alpha^2 + \beta
        \left(1 - \tan2\theta\right)}\right| \nonumber,
    \ee
whereas the width for $c_{\text{L}}$-type effects only is consequently given by
    \be
        \left|\Delta E^{\text{c}}(\text{FWHM})\right| = \left|\sqrt{\beta
        \left(1 + \tan2\theta\right)} - \sqrt{\beta \left(1 - \tan2\theta\right)}\right|.
    \ee
It is obvious that the sensible definition of $\Delta E^{\text{c}}(\text{FWHM})$ requires a 
scenario in which $\text{sgn}~\beta = +1$. In this case the ratio of the widths for combined 
$a_{\text{L}}$- and $c_{\text{L}}$-type effects and $c_{\text{L}}$-type effects satisfies
    \be
        \left|\frac{\Delta E^{\text{ac}}(\text{FWHM})}{\Delta E^{\text{c}}(\text{FWHM})}\right| \le 1.
    \ee
Therefore a combination of $a_{\text{L}}$ and $c_{\text{L}}$ effects leads to narrower resonances 
as compared to $c_{\text{L}}$-like effects only. This behavior is illustrated for the sines of the effective mixing angles in Fig. \ref{sina+1m+1}.

\section{Discussion}\label{discussion}

We begin our discussion by fixing a certain choice of signs for the matter-like parameter $a$ and
the shortcut-like (or Lorentz-violating) parameter $b$ and $\Delta m^2$.
We seek a combination which establishes a resonance in neutrinos, which will be matched with the MiniBooNE
excess around $E^{\text{res}}_{\text{MiniBooNE}} = 200~\text{MeV}$, and a antineutrino resonance to be found below the MiniBooNE value in order to match the value $E^{\text{res}}_{\text{LSND}} = 50~\text{MeV}$ of LSND. This can be 
accommodated for electron neutrinos in a flavor-aligned framework for example, i.e. a choice $\text{sgn}~a = +1$,
$\text{sgn}~b = -1$ and $\text{sgn}~\Delta m^2 = -1$ is required. We thus would explain both MiniBooNE
and LSND via resonant neutrino oscillations and since the antineutrino resonance lies in the kinematic region of
LSND we would not expect a signal in the antineutrino channel of MiniBooNE (which is actually not observed).
For the case of two resonances, namely $E^{-+}_{\text{res}} \equiv E^{\nu}_{\text{res}}$ and 
$E^{++}_{\text{res}} \equiv E^{\bar{\nu}}_{\text{res}}$, an order of magnitude estimate yields
   \be 
      \left|a\right| = 0.5 \times 10^{-8} \times 
      \left(\frac{\left|\Delta m^2 \cos2\theta\right|}{1~\text{eV}^2}\right) 
      \left(\frac{200~\text{MeV}}{E^{\nu}_{\text{res}}}\right) \left(\frac{E^{\nu}_{\text{res}}}{E^{\bar{\nu}}_{\text{res}}}-1\right)~\text{eV}
   \ee
and accordingly
   \be
      \left|b\right| = 10^{-16} \times \left(\frac{\left|\Delta m^2 \cos2\theta\right|}{1~\text{eV}^2}\right) 
      \left(\frac{200~\text{MeV}}{E^{\nu}_{\text{res}}}\right) 
      \left(\frac{50~\text{MeV}}{E^{\bar{\nu}}_{\text{res}}}\right),
   \ee
where clearly $E^{\nu}_{\text{res}} > E^{\bar{\nu}}_{\text{res}}$ has been assumed. The matter density $\rho$
for neutral matter composed of light elements, as typically encountered in earth-bound neutrino oscillation experiments, on the other hand can be written as
   \be
      \rho \simeq 10^{14} \times \left(\frac{\left|a\right|}{1~\text{eV}}\right)~\frac{\text{g}}{\text{cm}^3}.
   \ee
It is obvious that standard matter effects only cannot account for the resonance structures in our model given the estimate on the CPT coefficients introduced above; they would be too weak by five orders of magnitude as matter
densities on earth are typically $\rho \approx 1~\text{g}/\text{cm}^3$. An explanation of LSND and MiniBooNE within
this scenario would thus require non-standard matter effects. Such non-standard matter effects arise, e.g. in the
model proposed in Ref. \cite{Nelson:2007yq}, where neutrino matter interactions are mediated
by exchange of a light $(B-L)$ gauge boson. Another possibility would be CPT-violating neutrino oscillations with non-vanishing $a$-type coefficients and a coefficient $b$ which has the same order of magnitude as predicted by models with
extra-dimensional shortcuts but the opposite sign (i.e. a generic Lorentz-violating term rather than a potential
generated by sterile neutrinos propagating through the extra-dimensional bulk). The same predictions would be obtained for the case of flavor-misalignment if one chooses $\text{sgn}~a = -1$, $\text{sgn}~b = +1$ and $\text{sgn}~\Delta m^2 = +1$ for electron neutrinos. This case 
suggests extra-dimensional shortcuts in connection with generic CPT-violating $a$-type coefficients mimicking matter effects as a possible explanation for the observed resonances in both LSND and MiniBooNE since the order of magnitude
estimate for $b$ nicely matches the one given earlier in Ref. \cite{Pas:2005rb}.
\par
However, this is not the only possible way to accommodate data in our model. It is also conceivable that there
only exists a resonance in the neutrino channel, as observed by MiniBooNE, and that LSND should be explained by
non-resonant neutrino oscillations \cite{Louis}. This could be realized in a framework with $\text{sgn}~a = +1$,
$\text{sgn}~b = -1$ and $\text{sgn}~\Delta m^2 = +1$ for flavor-alignment or $\text{sgn}~a = -1$,
$\text{sgn}~b = +1$ and $\text{sgn}~\Delta m^2 = -1$ for flavor-misalignment for electron neutrinos in each case.
Resonances for antineutrinos in those scenarios would be inexistent. This yields
   \be
      \left|a\right|^2 > 4 \times 10^{-16} \times 
      \left(\frac{\left|\Delta m^2 \cos2\theta\right|}{1~\text{eV}^2}\right)
      \left(\frac{\left|b\right|}{10^{-16}}\right)~\text{eV}^2
   \ee
assuming again a typical order of magnitude for $b$ motivated by extra-dimensional shortcut scenarios; thus 
this condition does not supplement the discussion much.
\par
A third way of confronting our model with the data at hand is the following hint:
The latest MiniBooNE data in the antineutrino channel reveal a slight excess of events for
an energy around 600 MeV. This finding could hint towards an explanation of the
MiniBooNE anomaly with a resonance energy of 200 MeV for the neutrino mode and a resonance energy of 600 MeV 
for the antineutrino mode and could be accommodated
in a shortcut-like scenario $\text{sgn}~b = +1$ with a mass hierarchy of $\text{sgn}~\Delta m^2 = +1$, e.g. for
flavor-alignment in the case of electron neutrinos,
where the LSND anomaly is due to non-resonant oscillations. Upcoming data in the MiniBooNE antineutrino mode will show whether the enhancement in the 475 -- 675 MeV energy range increases or decreases in significance \cite{Louis}.
For this case the $a$ and $b$ coefficients obey
   \be 
      \left|a\right| = 0.5 \times 10^{-8} \times 
      \left(\frac{\left|\Delta m^2 \cos2\theta\right|}{1~\text{eV}^2}\right) 
      \left(\frac{200~\text{MeV}}{E^{\nu}_{\text{res}}}\right) \left(1 - \frac{E^{\nu}_{\text{res}}}{E^{\bar{\nu}}_{\text{res}}}\right)~\text{eV}
   \ee
as well as 
\be
      \left|b\right| \simeq 0.8 \times 10^{-17} \times \left(\frac{\left|\Delta m^2 \cos2\theta\right|}{1~\text{eV}^2}\right) 
      \left(\frac{200~\text{MeV}}{E^{\nu}_{\text{res}}}\right) 
      \left(\frac{600~\text{MeV}}{E^{\bar{\nu}}_{\text{res}}}\right),
   \ee
respectively. This scenario again advocates an explanation not via standard matter effects but rather via
generic CPT-violating coefficients amending the common neutrino oscillation picture in the active sector. The value for $b$ again nicely fits the result found earlier for extra-dimensional shortcut scenarios.

\section{Conclusions}\label{conclusion}

We have studied the interplay between altered dispersion relations in two-state active-sterile neutrino oscillations. It has been assumed that the active and the sterile sector exhibit different dispersion relations. Such altered dispersion relations for active and sterile neutrinos naturally arise in scenarios with CPT and Lorentz symmetry violating extensions of the standard model. We discussed standard matter effects and models with sterile neutrinos taking shortcuts through an extra-dimensional bulk as two distinct subsets of such scenarios. 
It has been shown that CPT violation effects in the active neutrino sector and altered dispersion relations with a different energy dependence for sterile neutrinos give rise to new resonance structures in neutrino oscillation phenomena.
The possible resonance structures depend on the available parameter space for altered dispersion relations potentials, the mass hierarchy of active and sterile neutrinos as well as the particle or antiparticle nature of neutrinos. We have shown that including CPT-violating effects for active neutrinos
in a framework with altered dispersion relations in the sterile sector (as suggested by extra-dimensional models) enhances the possible resonance features and leads to narrower resonance widths in comparison with altered dispersion relations for sterile neutrinos in vacuo.
\par
Last but not least we analyzed the consequences of such scenarios in the context of the LSND and MiniBooNE oscillation anomalies and showed
that in an approach using potentials generated by CPT symmetry violation the LSND and MiniBooNE data may be explained via non-standard terms in the neutrino oscillation Hamiltonian.
Moreover it is possible that LSND is explained by non-resonant oscillation phenomena. This approach allows for an explanation of the observed MiniBooNE anomaly using CPT-violating effects for active neutrinos and extra-dimensional shortcuts, if the slight excess of events in the oscillation data around 600 MeV in the antineutrino channel increases in significance. However, even if this should prove to be not the case it is still possible in our model to
accommodate experimental findings which only reveal a resonance-like excess of events in the neutrino channel, whereas
LSND could then be explained by non-resonant oscillations.
While matter effects due to standard model weak interactions turn out to be negligible at sub-GeV energies of accelerator neutrino experiments, new interactions in standard model extensions could give rise to non-standard
CPT-violating effects affecting the analysis of the LSND and MiniBooNE anomalies. Such non-standard effects could reduce the widths of resonance peaks and shift the resonance energies of neutrinos with respect to the resonance
energies of antineutrinos, thus improving data fits for altered dispersion relation solutions to the LSND and MiniBooNE anomalies.

%\acknowledgments
%We thank W.C. Louis for useful discussions and an important comment on the
%possibility to explain LSND via non-resonant oscillations.

\vspace{1.5cm}\noindent
{\bf Note added:}
\par
When version 2 of this paper had been prepared for resubmission, the MiniBooNE collaboration announced an 
evidence for an electron-like excess in the anti-neutrino data above 475 MeV, being consistent with the expectation from a 2-neutrino fit to LSND \cite{Water}.
As there was no such finding in the neutrino data sample,
this result adds to the interestingness of the scenarios discussed in this 
paper.

\end{document}